\documentstyle[12pt,twoside,psfig,fleqn,espcrc1]{article}

\newcommand{\AmS}{{\protect\the\textfont2
  A\kern-.1667em\lower.5ex\hbox{M}\kern-.125emS}}

\title{Nuclear Astrophysics}

\author{K. Langanke
\address{Institute for Physics and Astronomy, University of
Aarhus, 
DK-8000 Aarhus C, Denmark}
        \thanks{The work was partly supported by the Danish Research
Council.}}

\begin{document}
\maketitle

\begin{abstract}
The manuscript reviews progress achieved in recent years in various
aspects of nuclear astrophysics, including stellar nucleosynthesis,
nuclear aspects of supernova collapse and explosion, neutrino-induced
reactions and their possible role in the supernova mechanism and 
nucleosynthesis, explosive hydrogen burning in binary systems, and
finally the observation of $\gamma$-rays from supernova remnants.
\end{abstract}

\section{Introduction}

In recent years nuclear astrophysics has grown into one of the major
subfields in nuclear physics and has become motivation for many on-going
and future developments worldwide. This is partly related to the fact
that the interest within nuclear astrophysics has also shifted in the
last years. Has the focus been in the past on the understanding of
hydrostatic burning in stars and related stellar evolution involving
mainly nuclear physics along the valley of stability (for a
delightful reference the reader is refered to Willy Fowler's Nobel
lecture \cite{Fowler84}), interest has shifted to astrophysical sites
and scenarios which require knowledge of nuclear processes and properties
at the two extreme sides of the nuclear chart, very proton- and
neutronrich nuclei. Clearly progress in these domains and hence more
reliable simulations of the astrophysical scenarios is expected from
radioactive ion beam (RIB) facilities and this is exactly the reason why the
(nuclear) astrophysics community so strongly supports the RIB iniatives
worldwide.

Important progress has been achieved in many aspects of nuclear
astrophysics.
Due to new observational tools (e.g. Hubble Space Telescope, COBE,
ROSAT, Superkamiokande) astronomical observations of the universe and
its contents cover now the different wavelengths of the
electromagnetic spectrum and detect particles and with increasing
importance also $\gamma$-rays which allow to draw conclusions about the
nuclear processes involved. Besides the various RIB iniatives, improvements
in laboratory nuclear astrophysics has also been made possible by new
detector developments as well as by moving underground. For example, the
LUNA collaboration, working in the Gran Sasso underground laboratory,
succeeded at measuring, for the first time, one of the reactions in the
pp-chain at those energies at which it proceeds most effectively in our
sun. Finally, progress in computer hard- and software going hand in hand
with improved nuclear and astrophysical models has contributed to more
realistic and better  simulations and understanding of various
astrophysical scenarios, but have also raised new questions which will
keep the field exciting and vivid in the coming years.

This review will attempt to present some of the highlights of the recent
years, but given the width of the field it cannot be complete
(for references to subjects not discussed here see the NUPECC report on
Nuclear and Particle Astrophysics \cite{Nupecc}). As a
red thread we will focus mainly on the nuclear physics related to the
understanding of type II supernovae including relevant nuclear input for
the collapse, the explosion mechanism and related nucleosynthesis
processes. Special attention will be paid
to the role played by neutrinos. Finally we will briefly discuss
explosive hydrogen burning in novae and x-ray bursters. But at the
beginning we like to report about the first successful studies of the
chemical evolution of a galaxy resembling our Milky Way.

\section{Galactical chemical evolution}

The general ideas of the synthesis of elements in the universe has
been derived now more than 40 years ago by Burbidge, Burbidge, Fowler
and Hoyle \cite{BBFH} and, independently, by Cameron \cite{Cameron}.
Due to this picture, the light elements (mainly hydrogen and helium)
have been made during the Big Bang, while the breeding places for most
of the other elements are stars. The stars generate the energy, which
allows them to stabilize and shine for millions of years and longer, by
transmuting nuclear species, thus forming new elements. These processes
occur inside the star, but are eventually released if the star for
example is massive enough and finally explodes in a type II supernova.
The freshly bred nuclear material is mixed into the interstellar medium
(ISM)
and can thus become part of the initial abundance composition for a new
star to be formed. Thus the galactical chemical evolution represents 
a `cosmic cycle', and the modellation of the observed solar abundances
(e.g. \cite{Pagel})
requires the simulations of the formation of a galaxy and of the stellar mass
distribution, birth rates,
evolution 
and lifetimes. Importantly one has to calculate the abundances 
produced by a star of a given mass and the amount and composition of
matter ejected into the ISM by the star's final type II supernova
explosion. Finally contributions of type Ia supernovae have to be added
which involve the formation and evolution of binary systems composed of
a giant star with a hydrogen envelope and an accreting white dwarf.

Despite its complexity, rather consistent studies of the galactical
chemical evolution have been performed by Woosley and collaborators
\cite{Timmes} and by Nomoto, Thielemann and collaborators
\cite{Tsujimoto}. Although the simulations involve still a few model
assumptions (from the nuclear input, the rate of the important 
$^{12}$C($\alpha,\gamma$)$^{16}$O reaction is still too uncertain
\cite{Buchmann}), excellent agreement is obtained 
with solar abundances \cite{Grevesse}
for 76 isotopes from hydrogen to zinc, when the calculation is sampled
at a time $4.55 \cdot 10^9$ years ago at a distance of 8.5 kpc from the
galaxy center (corresponding to birth time and position of our sun in
the Milky Way). As can be seen in Fig. 1, most of the abundances agree
within a factor of 2.

\begin{figure}[htb]
\begin{center}
  \leavevmode
  \psfig{figure=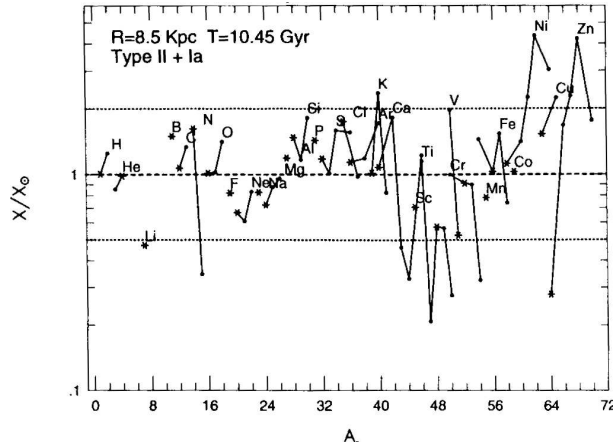,width=0.5\textwidth} 
\caption{
Ratio of the calculated and observed solar abundances for stable isotopes from
hydrogen to zinc. The dotted lines mark deviations by a factor of 2
between calculation
and observation. (from \protect\cite{Timmes})
}
\label{fig:galactic}
\end{center}
\end{figure}

It should be mentioned that specific nuclei appear to be almost entirely
(e.g. $^{11}$B, $^{19}$F) or in a large fraction (e.g. $^{10}$B,
$^{15}$N) made by neutrino nucleosynthesis \cite{Woosley90}. These
nuclei are the product of reaction sequences induced by neutral current
$(\nu,\nu')$ reactions on very abundant nuclei like $^{12}$C, $^{16}$O
and $^{20}$Ne,
when the flux of neutrinos generated by cooling of the neutron
star passes through the overlying shells of heavy elements.

\section{Nuclear physics input in supernovae}

At the end of hydrostatic burning, a massive star consists of concentric
shells that are the remnants of its previous burning phases (hydrogen,
helium, carbon, neon, oxygen, silicon).
Iron is the final stage of nuclear fusion in hydrostatic burning, 
as the synthesis of any heavier
element from lighter elements does not release energy; rather, energy
must be used up. If the iron core, formed in the center of the massive
star, exceeds the Chandrasekhar mass limit of about 1.44 solar masses,
electron degeneracy pressure cannot longer stabilize the core and it
collapses starting what is called a type II supernova. In its aftermath
the star explodes and parts of the iron core and the outer shells are
ejected into the ISM. Although this general picture has been confirmed
by the various observations from supernova SN1987a, simulations of the
core collapse and the explosion are still far from being completely 
understood and robustly modelled. In the following subsections we will
briefly review some recent progress, related to nuclear input, which
might help for more reliable supernova simulations.

\subsection{Weak interaction rates for the core collapse}

As pointed out by Bethe {\it et al.} 
\cite{Bethe78,Bethe90} the
collapse is very sensitive to the entropy and to the number of leptons
per baryon, $Y_e$. In turn these two quantities are mainly determined
by weak interaction processes, electron capture and $\beta$ decay. 
First, in the early stage 
of the collapse $Y_e$ is
reduced as electrons are captured by  Fe-peak nuclei. 
This reduces the electron pressure, thus accelerating the collapse, and shifts 
the distribution of nuclei present in the core to more neutron-rich
material. Second, many of the nuclei present can also $\beta$ decay.
While this process is quite unimportant compared to electron capture for
initial $Y_e$ values around 0.5, it becomes increasingly competative for
neutron-rich nuclei due to an increase in phase space related to larger
$Q_\beta$ values.

Knowing the importance of the weak interaction process, Fuller {\it et
al.} (usually called FFN)
have systematically estimated the rates for nuclei in the mass
range $A=45-60$ putting special emphasis on the importance of electron
capture to
the Gamow-Teller (GT) giant resonance \cite{FFN}. 
Another important idea in FFN was to recognize the role played by the GT
resonance in $\beta$ decay. Other than in the laboratory, $\beta$ decay
under stellar conditions are significantly increased due to thermal
population of the GT back resonance in the parent nucleus (the GT back
resonance are the states reached by the strong GT transitions in the
inverse process (electron capture) built on the ground and excited
states \cite{FFN,Aufderheide94})
allowing for a transition with a large nuclear matrix element and
increased phase space. Indeed, Fuller et al. concluded that the
$\beta$ decay rates under collapse conditions are dominated by the decay
of the back resonance.

The GT contribution to the electron capture and $\beta$ decay rates
has been parametrized by FFN on the basis of the independent particle
model. To complete the FFN rate estimate, the GT contribution has been
supplemented by a contribution simulating low-lying transitions.
Recently the FFN rates have been updated and extended to heavier nuclei
by Aufderheide {\it et al.} \cite{Aufderheide94}. 

In recent years, however, the parametrization of the GT contribution, as
adopted in \cite{FFN,Aufderheide94}, has become questionable when
experimental information about the GT distribution in nuclei became
available.
These  data clearly indicate that the GT
strength is not only quenched, 
but also fragmented over several
states at modest excitation energies in the daughter nucleus
\cite{gtdata1,gtdata2,gtdata3,gtdata4,gtdata5}. Thus the need
for an improved theoretical description has soon been realized
\cite{Aufderheide91,Aufderheide93a,Aufderheide93b}, but it became also
apparent that a reliable reproduction of the GT distribution in nuclei
requires large shell model calculations which account for all correlations
among the valence nucleons in a major oscillator shell.
Such $0 \hbar \omega$ shell model calculations are now possible and they
come in two varieties: 
large-scale
diagonalization approaches \cite{Caurier98} 
and shell model Monte Carlo (SMMC) techniques \cite{Johnson92,Physrep}.
The latter can treat the even larger model spaces, but has limitations
in its applicability to odd-A and odd-odd nuclei at low temperatures, 
which does not apply to the former.
More importantly the diagonalization approach allows for detailed
spectroscopy, while the SMMC model yields only an ``averaged'' GT
strength distribution which introduces some inaccuracy into the
calculation of the capture and decay rates. 

Core collapse electron capture and $\beta$ decay rates have now been
calculated for the relevant nuclei 
in the mass range $A=46-65$ \cite{Langanke99}. The
calculations have been performed by
shell model diagonalization 
in model spaces large enough (involving typically 10 million or more
configurations)
to guarantee that the GT
strength distribution is virtually converged.
Adopting a recently developed, 
improved version of the KB3 interaction \cite{ni56} these calculations
reproduce all measured GT strength
distributions very well 
(after scaling by the universal factor, $(0.74)^2$
\cite{Wildenthal,Langanke95,Martinez96})
and  describe the experimental level
spectrum of the nuclei involved quite accurately
\cite{Martinez99}.  
Fig. 2 compares the measured and calculated GT distributions for several
nuclei
in the mass range $A=50-64$. 
Furthermore modern large-scale shell model
calculations also agree with measured half-lifes very well. Thus for the
first time one has a tool in hand which allows for a reliable
calculation of presupernova electron capture and $\beta$ decay rates. 

\begin{figure}[htb]
\begin{center}
  \leavevmode
  \psfig{figure=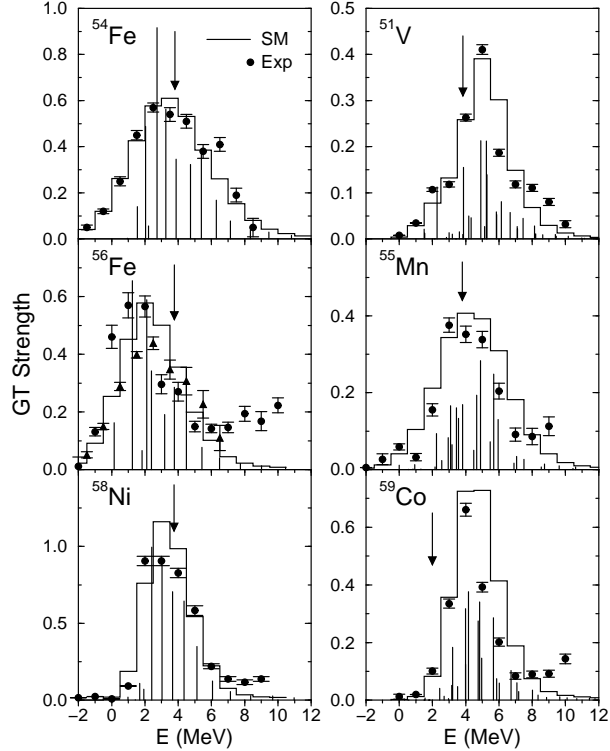,width=0.5\textwidth} 
\caption{
Comparison of the shell model GT strength distribution (histogram)
with data \protect\cite{gtdata1,gtdata2,gtdata3,gtdata4,gtdata5} 
for selected even-even 
(right) and odd-A nuclei (left). For the comparison the calculated 
discrete spectrum has
been folded with the experimental resolution. The positions of the GT
centroid assumed in the FFN parametrization are shown by arrows.
}
\label{fig:gtstrength}
\end{center}
\end{figure}

Data and shell model calculations indicate  systematic differences in the
location of the main GT resonance strength compared to the
parametrization of FFN \cite{Dean98,Langanke99}.  
In capture on even-even nuclei the GT strength
resides at lower excitation energies in the daughter than assumed by
FFN, while in odd-A and odd-odd nuclei the GT strength is centered at higher
excitation energies. As a consequence, the shell model electron capture
rates on odd-A and odd-odd nuclei are significantly (by an order of
magnitude or more) smaller than the compiled rates, while the capture rates on
even-even nuclei are approximately the same, as FFN had often intuitively
compensated the smaller GT contribution (compared with the shell model)
by the empirically added low-lying transition strength. 
Which consequences do the misplacement of the GT centroids have for the
competing $\beta$ decays? In odd-A and even-even nuclei (the daughters
of electron capture on odd-odd nuclei), experimental 
data and shell model studies
place the back-resonance at higher excitation energies than assumed by
FFN and Aufderheide et al. \cite{Aufderheide94}. Correspondingly, its
population becomes less likely at the temperatures available during the
early stage of the collapse ($T_9 \approx 5$, where $T_9$ measures the
temperature in $10^9$ K) and hence the contribution of the back-resonance
to the $\beta$ decay rates for even-even and odd-A nuclei decreases. 
In contrast, the shell model $\beta$ decay rate for odd-odd nuclei
often are slightly larger than the FFN rates, as
for these nuclei, all available data, stemming from
(n,p) reaction cross section measurements on even-even nuclei like
$^{54,56,58}$Fe or $^{58,60,62,64}$Ni, and all shell model calculations
indicate that the back-resonance resides actually at lower excitation
energies than previously parametrized.

What might the revised electron capture and 
$\beta$ decay rates mean for the core collapse?
One can investigate this question by studying the change of the
electron-to-baryon ratio, ${\dot Y}_e$, along a stellar trajectory
as given in Ref.
\cite{Aufderheide94}.
As can be seen in Fig. 3,
the shell model rates reduce 
${\dot Y}_e^{ec}$ significantly compared with the FFN; 
by more than an order of magnitude
for $Y_e<0.47$.
This is due to the fact, that,
except for $^{56}$Ni, all shell model electron capture rates
are smaller than the recommendations given in the FFN  
compilations \cite{FFN}.
The shell model $\beta$ decay rates also reduce 
${\dot Y}_e^{\beta}$, however, by a smaller amount than for electron
capture. This is mainly caused by the fact that
the shell model $\beta$
decay rates of odd-odd nuclei are about the same as the FFN rates or
even slightly larger, for reasons discussed above.  
The important feature in Fig. 3 is that $\beta$ decay rates are larger
than the electron capture rates for $Y_e=0.42-0.455$, which is also
already true for the FFN rates. This, however, 
had apparently not been considered
in collapse calculations yet which used quite outdated $\beta$ decay
rates \cite{Aufderheide94a}. 

The consequences of the shell model rates on the core collapse has to be
explored by detailed and consistent calculations. One expects that 
in the early stage of the collapse the stellar trajectory is, for a
given density, at a higher temperature, as, due to the slower electron
capture rates, the star radiates less energy away in form of neutrinos
until $Y_e=0.455$ is reached, and the $\beta$ decay rates become larger
than the electron rates. This might lead to cooler cores and
larger $Y_e$ values at the formation of the homologuous core, as pointed
out in \cite{Aufderheide94a}.

\begin{figure}[htb]
\begin{center}
  \leavevmode
  \psfig{figure=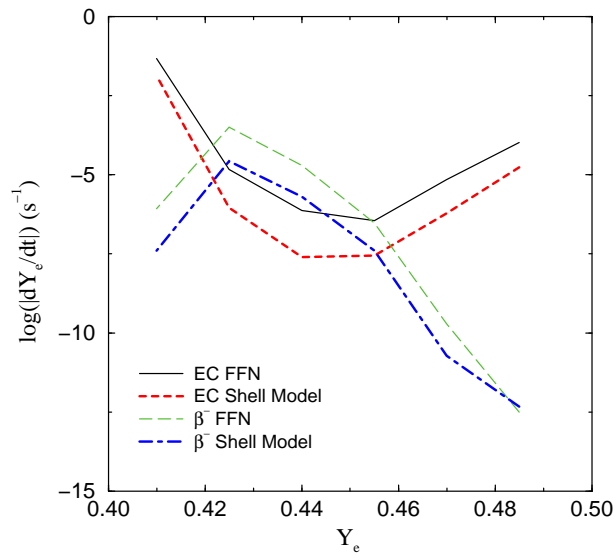,width=0.5\textwidth} 
\caption{
Change in the total electron capture and $\beta$ decay rates, 
${\dot Y}_e^{ec}$ and ${\dot Y}_e^{\beta}$, respectively. The shell
model results are compared with the FFN results \protect\cite{FFN}
along the same stellar
trajectory as in Fig. 14 of Ref. \protect\cite{Aufderheide94}.
}
\label{fig:evol}
\end{center}
\end{figure}

\section{Towards a robust supernova explosion}

Electron capture, $\beta$ decay and photodisintegration cost the core
energy and reduce its electron density. As a consequence, the collapse
is accelerated. An important change in the physics of the collapse
occurs, if the density reaches $\rho_{\rm trap} \approx 4 \cdot 10^{11}$
g/cm$^3$. Then neutrinos are essentially trapped in the core, as their
diffusion time (due to coherent elastic scattering on nuclei) becomes
larger than the collapse time. After neutrino trapping, no energy is
carried away from the core. Shortly after (at $\rho \approx 10^{12}$
g/cm$^3$), neutrinos are  thermalized by inelastic scattering on
electrons. Then all reactions are in equilibrium, including the weak
processes discussed above. The degeneracy of the (trapped) neutrino Fermi
gas hinders a complete neutronization. As a consequence, $Y_e$ remains
rather large through the collapse ($Y_e=0.35-0.38$ \cite{Bethe90},
possible consequences due the modified weak interaction rates have to be
explored). To balance the charge, the number of protons must therefore
also be large and this can only be achieved in heavy nuclei. The
collapse has a rather large order and the entropy stays small during the
collapse \cite{Bethe78}. 

After neutrino trapping, the collapse proceeds homologously
\cite{Goldreich}, until nuclear densities 
($\rho_N\approx10^{14}$ g/cm$^3$)
are reached. As nuclear matter has a finite compressibility, the
homologous core decelerates and bounces in response to the increased
nuclear matter pressure; this eventually drives an outgoing shock wave
into the outer core; i.e. the envelope of the iron core outside the
homologous core,  
which in the meantime has continued to fall inwards at
supersonic speed. The core bounce with the formation of a shock wave is
believed to be the mechanism that triggers a supernova explosion, but
several ingredients of this physically appealing picture and the actual
mechanism of a supernova explosion are still uncertain and
controversial. 
If the shock wave is strong enough not only to stop
the collapse, but also to explode the outer burning shells of the star, 
one speaks about the `prompt mechanism'. However, it appears as if the
energy available to the shock is not sufficient, and the shock will
store its energy in the outer core, for example, by excitation of
nuclei.

\begin{figure}[htb]
\begin{center}
  \leavevmode
  \psfig{figure=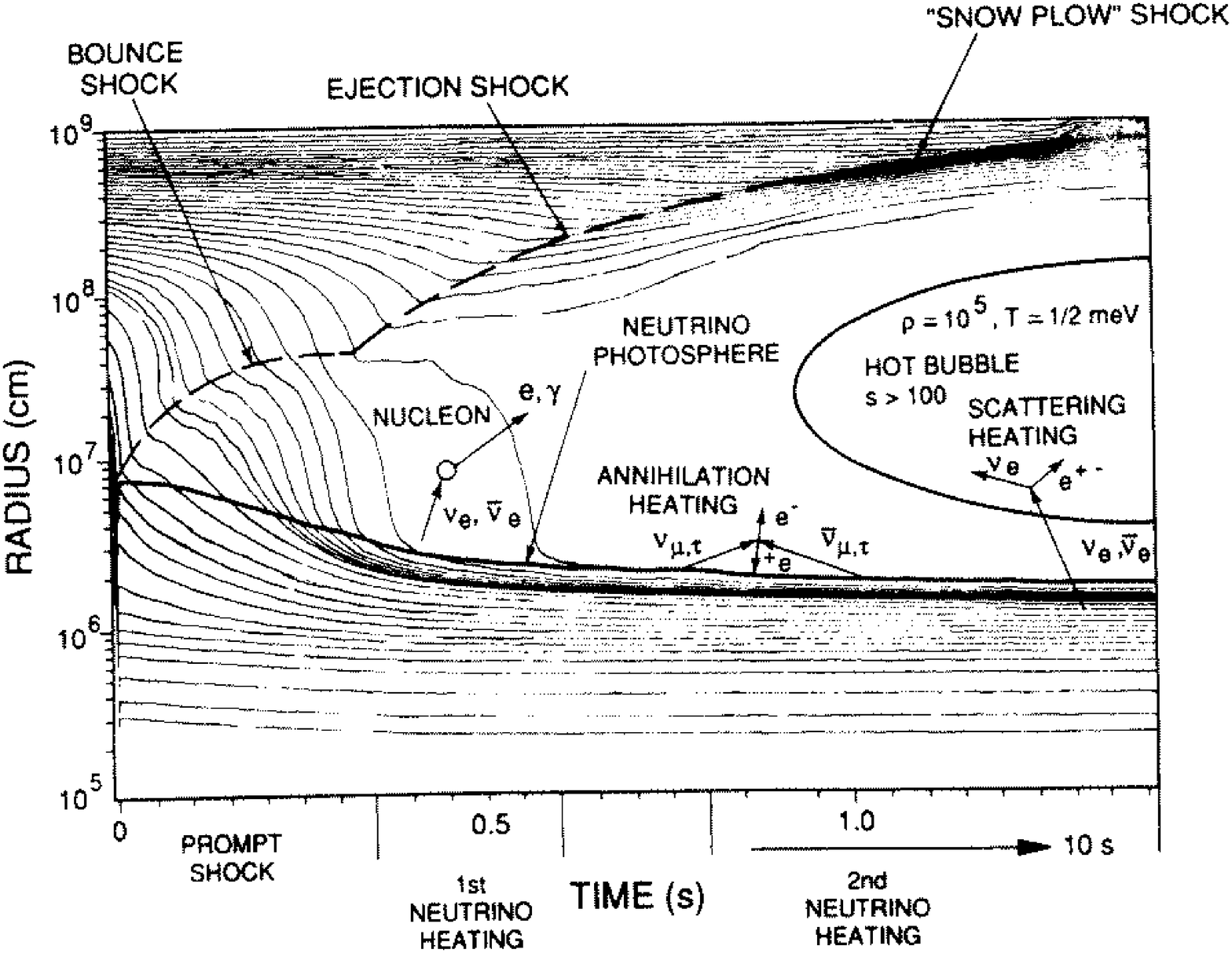,width=0.7\textwidth} 
\caption{
Mass trajectories during a core collapse leading to the discovery of the
delayed explosion mechanism by Wilson. 
The bounce occurs at time $t=0$. The shock travels outwards for a short
time, gets stalled and is later revived by the neutrinos generated due
to cooling of the nascent neutron star. Above the neutron star forms the
radiation bubble which might be the site for the nuclear r-process.
(from \protect\cite{Colgate}).
}
\label{fig:masstraj}
\end{center}
\end{figure}

After the supernova has exploded, a compact remnant with a gravitational
mass of order one solar mass is left behind; this mass is slightly
increased to 1.3-1.5 solar masses in the first second after the bounce
by accretion. The remnant is very lepton rich (electrons and
neutrinos), the latter being trapped  as their mean
free paths in the dense matter is significantly shorter than the radius
of the neutron star. It takes a fraction of a second \cite{Burrows90}
for the trapped
neutrinos to diffuse out, giving most of their energy to the neutron star
during that process and heating it up. The cooling of the protoneutron
star then proceeds by pair production of neutrinos of all three
generations which diffuse out. After several tens of seconds the star
becomes transparent to neutrinos and the neutrino luminosity drops
significantly \cite{Burrows88}. 
In the `delayed mechanism', the shock wave can be revived by these
outward diffusing neutrinos, which 
carry most of the energy set free in the gravitational collapse of the core.
The delayed mechanism has been discovered by Wilson \cite{Wilson}; the
mass trajectories of his computer simulation are shown as function of
time after the bounce in Fig. 4.

Although the details of the neutrino distributions leaving the
protoneutron star are still subject of research, it is generally
accepted that there is a temperature hierarchy between the various
neutrino types introduced by the charged current reactions with the
surrounding neutron-rich matter. Thus, $\mu$ and $\tau$ neutrinos and
their antiparticles (usually combiningly refered to as $\nu_x$
neutrinos) have the distribution with 
the highest temperature ($T=8$ MeV if one accepts a Fermi-Dirac
distribution with zero chemical potential \cite{Woosley90}). All $\nu_x$
neutrinos have the same distribution and thus potential oscillations
between these neutrino types as suggested by the Superkamiokande results
on atmospheric neutrinos \cite{Fukuda} are not important for the topics
discussed below. As the $\nu_e$ and ${\bar \nu_e}$ neutrinos interact
with the neutron-rich matter via $\nu_e + n \rightarrow p + e^-$ and
${\bar \nu}_e + p \rightarrow n + e^+$, the 
${\bar \nu}_e$ neutrinos have a higher temperature ($T\approx
5.6$ MeV) than the  $\nu_e$ neutrinos 
($T=4$ MeV). It is useful for the following discussions to
note that these temperatures correspond to average neutrino energies of
${\bar E_\nu}=25$ MeV for $\nu_x$ neutrinos, while
${\bar E_\nu}=16$ MeV and 11 MeV for ${\bar \nu}_e$ 
and $\nu_e$ neutrinos.

In the delayed supernova mechanism, neutrinos deposit energy in the
layers between the nascent neutron star and the stalled prompt shock.
This lasts for a few 100 ms, and requires about $1\%$ of the neutrino energy
to be converted into nuclear kinetic energy. The energy deposition increases the
pressure behing the shock and the respective layers begin to expand,
leaving between shock front and neutron star surface a region of low
density, but rather high temperature. This region is called the `hot
neutrino bubble' and, as we will discuss below, might be the site of the
nuclear r-process. The persistent energy input by neutrinos 
keeps the pressure high in this region and drives the shock outwards
again, eventually leading to a supernova explosion.

It has been found that  the delayed supernova mechanism is quite
sensitive to physics details deciding about success or failure in the
simulation of the explosion. Very recently, two quite distinct
improvements have been proposed which should make
the explosion mechanism more robust, as they increase the efficiency
of the energy
transport to the stalled shock.

\subsection{Convection}

Two-dimensional \cite{Burrows92} and three-dimensional
\cite{Mueller97} hydrodynamic simulations have recently become
possible and have clearly demonstrated the existence of convective
instabilities inside the protoneutron star. The convective velocities
can even reach the local sound speed, generating strong pressure waves
by the convective flow \cite{Mueller98}. Near the protoneutron star
surface the convective mixing is accompanied by an increase of the
neutrino luminosities during the early phase of the explosion; the
effect can be seen by comparing models B1 and B2 in Fig. 5. Due to the
rather violent convection, neutrinos are transported out of the dense
interior of the protoneutron star much faster than by diffusion,
making the revival of the shock more efficient.

\begin{figure}[htb]
\begin{center}
  \leavevmode
  \psfig{figure=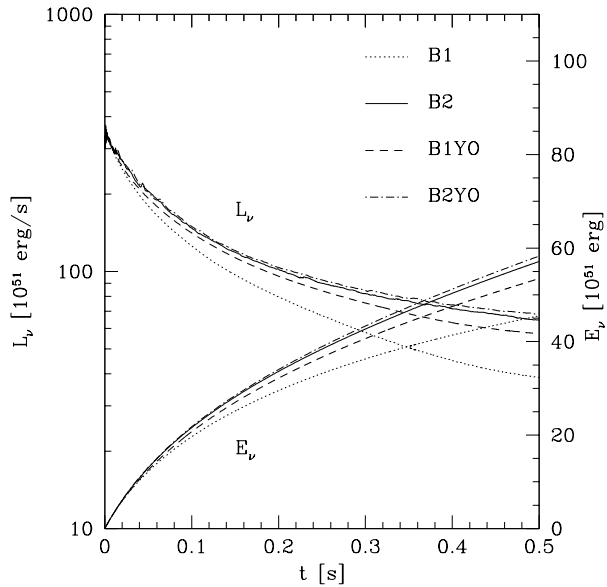,width=0.5\textwidth} 
\caption{
Total surface neutrino luminosity $L_\nu$ and integrated energy emission
$E_\nu$ as functions of time for models without (B1) and with (B2)
convection and with standard neutrino opacities (B1YO)  and reduced
neutrino opacities (B2YO)
(courtesy of H.-T. Janka,
\protect\cite{Janka}).
}
\label{fig:htj}
\end{center}
\end{figure}

\subsection{Neutrino opacities in dense matter}

Numerical simulations have also found convective instabilities in the
`hot neutrino bubble' region \cite{Herant,Burrows95,Janka96,Shimizu93}.
However, a realistic description for this region requires an adequate
treatment of the neutrino transport and opacities in dense matter. For a
long time, this subject has received rather little attention, but within
the last year it has evolved into a vividly studied subject.

To describe charged- and neutral-current reactions of neutrinos in
nuclear matter, one has to know the response of the system usually
characterized by dynamic form factors $S(q_0,q)$, where $q_0$
is the energy transfer to the baryons and $q$ the momentum transfer.
At the mean field level, the function $S$ can be evaluated exactly if
the single particle dispersion relation is known \cite{Reddy98}.
Compared to the case of a non-interacting system, strong interaction
corrections are incorporated by adopting an effective mass parameter,
which, like the single particle potentials, are chosen as density
dependent \cite{Reddy98}.

However, nucleon-nucleon correlations introduced by the residual
interaction (for which, for example, a Migdal parametrization like in
Fermi-liquid theory can be adopted, potentially extended by a tensor
component 
\cite{Reddy99,Akmal98,Sawyer99}) 
turn out to be quite important \cite{Sawyer98,Reddy99}. 
For example, these effects can be
studied within the random phase approximation. Then, neutral current
reactions are sensitive to the density-density and spin-density response
functions of nuclear matter, while charged current reactions depend on
the isospin density and spin-isospin density response functions
\cite{Reddy99}. 

\begin{figure}[htb]
\begin{center}
  \leavevmode
  \psfig{figure=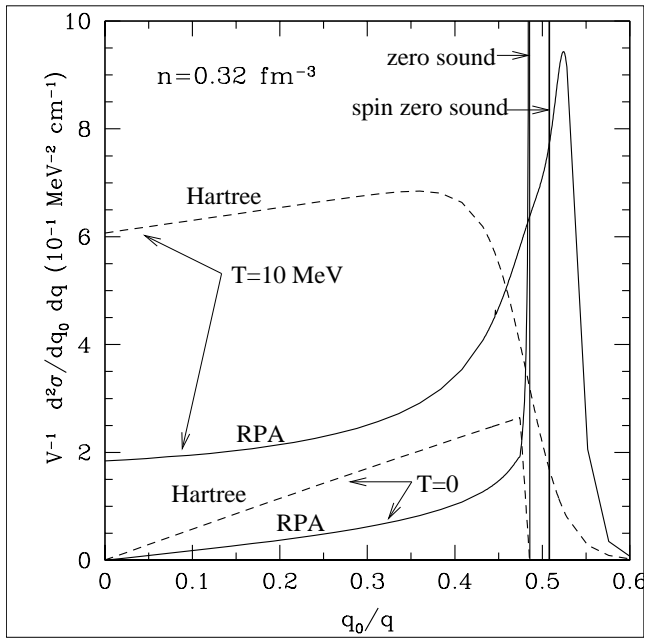,width=0.5\textwidth} 
\caption{
Neutrino scattering cross section as function of energy transfer $q_0$
in pure neutron matter for momentum transfer $q=10$ MeV and neutrino
energy $E_\nu=50$ MeV. (from \protect\cite{Reddy98a}) 
}
\label{fig:htj}
\end{center}
\end{figure}

For neutral-current reactions like neutrino-neutron scattering, it is
found that the repulsive interaction shifts strength to larger $q_0$
enhancing the collective excitations, while the cross sections are
significantly reduced at small energy transfers $q_0$, as is
demonstrated in Fig. 6 for pure neutron matter. The same effects are
observed in charged-current reactions, where the excitation of the
Gamow-Teller and giant dipole resonances take away strength from low
$q_0$. 

It appears that nucleon-nucleon correlations reduce the neutrino
opacities in dense nuclear matter considerably compared with the free
gas estimate \cite{Sawyer98,Reddy99}. This might imply shorter time
scales for the deleptonization and cooling of the protoneutron star and
a more efficient energy transport to the stalled shock (see also the
models B1YO and B2YO in Fig. 5).  

\section{R-process in the hot neutrino bubble}

Within the last few years the neutrino-driven wind model has been widely
discussed as the possible site of r-process nucleosynthesis 
\cite{Woosley94a,Witti}. Here it is assumed that the r-process occurs in the
layers heated by neutrino emission and evaporating from the hot
protoneutron star after core collapse. In this model
(e.g. \cite{Thielemann98}), a hot blob
of matter with entropy per baryon $S_b$ and 
electron-to-baryon ratio $Y_e$, initially consisting of neutrons, protons
and $\alpha$-particles in nuclear statistical equilibrium (NSE), expands
adiabatically and cools. Nucleons and nuclei combine to heavier nuclei,
with some neutrons and $\alpha$-particles remaining. Depending on
the value of $S_b$, the nuclei produced are in the iron group or, at
higher entropies, can have mass numbers $A=80-100$. These nuclei then
become the seeds and, together with the remaining neutrons, undergo an
r-process \cite{Thielemann}. 
In this model a successful r-process 
depends mainly on four parameters: 
the entropy per baryon $S_b$,
the dynamical timescale, the mass loss rate and the electron-to-baryon ratio
$Y_e$. All parameters depend on the neutrino luminosity and are determined
mostly by $\nu_e$ and ${\bar \nu_e}$ absorption on free 
nucleons. During a supernova explosion these parameters vary and the
r-process in the hot neutrino bubble becomes a dynamical and
time-dependent scenario. Woosley et al. \cite{Woosley94a}
have calculated the r-process
abundance for this site, adopting the parameters as given by Wilson's
successful supernova model (e.g. Fig. 4). The final abundances obtained
after integration over the duration of the r-process in the hot neutrino
bubble (several seconds) are shown in Fig. 7, showing quite satisfying
agreement between calculation and observation. 

\begin{figure}[htb]
\begin{center}
  \leavevmode
  \psfig{figure=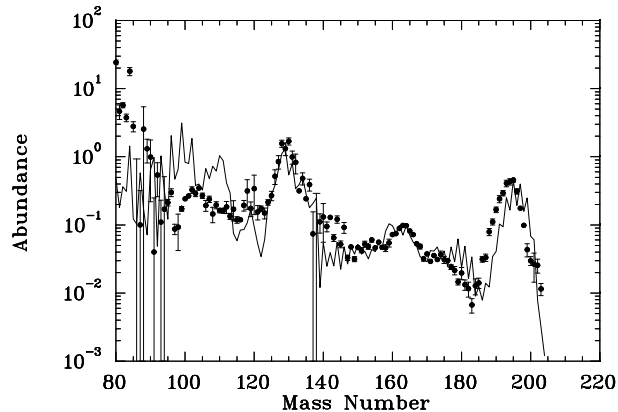,width=0.5\textwidth,angle=90} 
\caption{
R-process abundances in the hot neutrino bubble model compared to
observation (from \protect\cite{Woosley94a})
}
\label{fig:meyer}
\end{center}
\end{figure}

Neutrino-induced reactions can be important during and even after
the r-process. In the conventional picture \cite{Thielemann}
the nuclei are basically in 
$(n,\gamma)$/$(\gamma,n$) equilibrium during the r-process. The r-process path
is mainly determined by neutron separation energies and 
the timescale is essentially
set by the $\beta$-decays of the waiting-point nuclei
at the magic neutron numbers $N=50,82$ and 126. However, in the presence
of a strong neutrino flux, $\nu_e$-induced charged-current reactions
on the waiting-point nuclei might actually compete with $\beta$-decays
and speed-up the passage through the bottle-necks at the magic neutron
numbers \cite{Qian}.
It is found \cite{Qian}
that, for typical neutrino luminosities and spectra, $\nu_e$-capture rates
are of order $5$ s$^{-1}$ and thus
can be faster than competing $\beta$-decays for the slowest waiting-point
nuclei. Of course, quantitative conclusions can only be drawn from
detailed numerical simulations of the r-process. A first step towards
this goal has recently been made by Meyer et al. \cite{Meyer98}.

It is usually assumed that the r-process 
drops out of $(n,\gamma$)/($\gamma,n$) equilibrium in a sharp freeze-out.
The very neutron-rich matter, assembled during the r-process,
then decays back to the valley of stability by a sequence of $\beta$-decays. 
However, in the neutrino-driven wind scenario the r-process matter will still be
exposed to rather strong neutrino fluxes, even after freeze-out.
By both $\nu_e$-induced charged-current and $\nu_x$-induced
neutral-current reactions, neutrinos can inelastically interact
with r-process nuclei. In these
processes the final nucleus will be in an excited state and most likely
decay by the emission of one or several neutrons. Thus, this
{\it post-processing} of r-process matter after freeze-out might effect
the final r-process abundance. 
The neutrino post-processing effects depend 
on the neutrino-induced neutron knock-out cross sections,
which Qian {\it et al.} \cite{Qian} have calculated based on the
continuum random phase approximation and the statistical model, and
on the total neutrino
fluence through the r-process ejecta following freeze-out.

The dominant features of the observed r-process abundance distribution
are the peaks at $A \sim 130$ and 195, corresponding to the progenitor
nuclei with $N=82$ and 126 closed neutron shells. Haxton et al.
\cite{Qian} find that 8 nuclei, lying in the window $A=124-126$ and
183-187, are unusually sensitive to neutrino post-processing
\cite{Haxton}. These
nuclei sit in the valleys immediately below the abundance peaks which
can be readily filled by spallation off the abundant isotopes in the
peaks. To avoid overproduction of the nuclei in these abundance windows
one is able to place upper bounds on the fluence (${\cal F} \leq 0.045$ at
$A\sim 130$ and $\leq 0.030$ at $A\sim 195$, respectively). Furthermore,
it turns out that the observed abundance of the nuclei 
in the two abundance windows can be consistently reproduced by the same
fluence parameter (for an example see Fig. 8). This might be taken as evidence
suggesting that the r-process does occur in an intense neutrino fluence,
and thus that the interior region of a type II supernova is the site of
the r-process.

We like to stress that the neutrino-induced knock-out liberates about
3-5 neutrons from nuclei in the abundance peaks around $A=130$ and 195.
Thus this process cannot be able to fill the well-developed abundance
trough at $A\approx 115$ \cite{Kratz} where r-process simulations with
conventional mass formulae strongly underestimate the observed
abundances. This discrepancy might point to interesting nuclear structure
effects in very neutron-rich nuclei, related to shell quenching far from
stability \cite{Chen}.

\begin{figure}[htb]
\begin{center}
  \leavevmode
  \psfig{figure=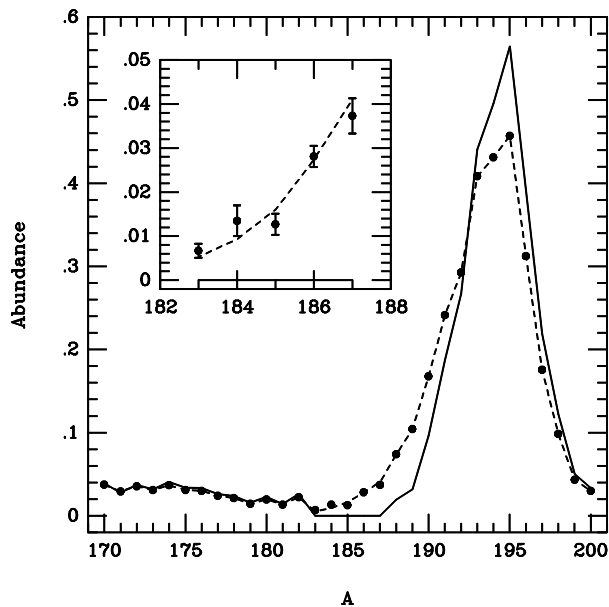,width=0.5\textwidth} 
\caption{
Effect of postprocessing by neutrino-induced reactions on the
r-process abundance. The unprocessed distribution (solid line) is
compared with the distribution after postprocessing (dashed line). A
constant fluence of
${\cal F}=0.015$ has been assumed which provides a best fit to the
observed abundances for
$A=183-87$   (see inset). The observed
abundances are plotted as filled circles with error bars. 
(from \protect\cite{Qian})
}
\label{fig:post}
\end{center}
\end{figure}

The open question currently is what kind of superpositions of entropies
the supernova neutrino-driven wind environment really provides.
In the supernova model used by Woosley et al \cite{Woosley94a} entropies
upto $S_b=300$ have been reached, but other models suggest that $S_b$ is a
factor of 3-5 smaller (e.g. \cite{Witti}). To understand the importance
of the entropy, one has to consider that the production of seed nuclei
has to go through the bottleneck of the 
3-body reaction $\alpha$+$\alpha$+n$\rightarrow \;
^{9}$Be at the start. Due to the low Q-value of this reaction ($Q=1.57$
MeV), a large entropy (or high photon number) drives this reaction in
equilibrium to the left, ensuring a rather small amount of $^9$Be. Since
all $^9$Be is basically transformed into seed nuclei, a high entropy
results in a small amount of seed nuclei and a large 
neutron-to-seed ratio $n/s$ 
\cite{Qian97}.

As a simple measure for a successful r-process one has to require the
presence of enough neutrons to synthesize nuclei in the r-process peak
around $A=200$ beginning from the seed nuclei ($A\le 100$). Thus, the
neutron-to-seed ratio $n/s$ has to reach at least 100 for a successful
r-process. 
Systematic studies by Hoffmann and collaborators 
\cite{Hofmann} and Freiburghaus {\it et
al.} \cite{Thielemann98}
have shown that
a successful r-process
requires either large entropies at the $Y_e$ values currently obtained
in supernova models, or smaller values for $Y_e$.

\section{Testing the supernova models}

Supernova SN1987A in the Large Magellanic Cloud confirmed the
fundamental ideas of the theoretical picture of type II supernovae.
Due to models, the supernova produces an appreciable amount of
$^{56,57}$Ni which decays within a few days to $^{56,57}$Co (the
supernova is not transparent at these early times and these decays are not
observed). The observed lightcurve of SN1987A then followed the
decay of $^{56}$Co (with a halflife of 77 days) and later
$^{57}$Co (271 days), exactly as expected \cite{Woosley89,Clayton91}, 
and a total amount of 0.075
solar masses of $^{56}$Co could be implied.  Now the
lightcurve is powered by the decay of $^{44}$Ti with a halflife of 60
years. The special role, played by the observation of $\gamma$ lines
originated from the decay of $^{44}$Ti, for the detection of past
supernovae will be reviewed below. However, before we like to briefly discuss
the detection of supernova neutrinos.

\subsection{Observation of $\nu_\mu$ and $\nu_\tau$ neutrinos 
in Superkamiokande}

In what is considered the birth of neutrino astrophysics and the most
outstanding single observation from supernova SN1987A, 19 neutrinos have
been detected 
by the Kamiokande
\cite{Kamio} and IMB \cite{IMB} water \v{C}erenkov detectors.
Although being only a few events, these neutrinos give evidence for the
collapse of the iron core of the evolved massive progenitor to a neutron
star, and even allow to derive constraints on neutrino properties.
It is generally assumed that these events originated
from the  
${\bar \nu}_e +p \rightarrow n+e^+$ reaction in water. 
The detection of 
$\nu_e$ and
$\nu_x$ neutrinos via the
$\nu+e\rightarrow \nu'+e'$ scattering 
or the
$^{16}$O($\nu_e,e^-$)$^{16}$F reaction
was strongly suppressed by the small effective cross sections
of these processes, although the $\nu_e$ induced signal can in principle
be separated by its angular distribution \cite{Haxton3}. 
Thus, SN1987A did not allow for a detailed test of the neutrino
distribution and, in particular, it gave no information about $\nu_x$
neutrinos, which as we discussed above, should decouple deepest in the
star. 
The observability of supernova 
neutrinos has significantly improved since
the Superkamiokande (SK) detector, with a threshold of 5 MeV,
is operational \cite{Totsuka}.
If we were lucky enough and a supernova occurs in our galaxy in the near
future, this detector is indeed  able to
supply the desired information about the neutrino distributions and temperature hierarchy, and can therefore test the change of neutrino opacities due
to the nuclear medium.

\begin{figure}[htb]
\begin{center}
  \leavevmode
  \psfig{figure=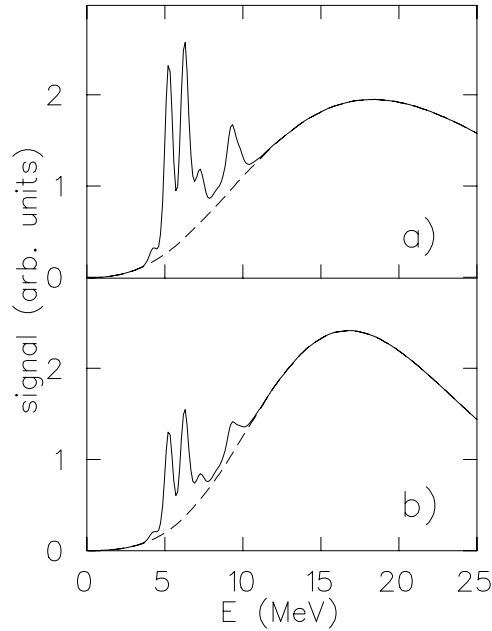,width=0.4\textwidth} 
\caption{
Signal expected from supernova neutrinos in a water Cerenkov detector
calculated for two different types of neutrino distributions (without
chemical potential (above) and with chemical potential $\mu=3T$ and
temperatures $T=6.26 $ MeV (for $\nu_x$) and $T=4$ MeV (for ${\bar
\nu}_e$)). The bulk of the signal stems from ${\bar \nu}_e$ neutrinos
reacting with protons, while the $\nu_x$ neutrinos induce the superimposed
signal at energies $E=5-10$ MeV. (from \protect\cite{Langanke})
}
\label{fig:kamio}
\end{center}
\end{figure}

Recently 
a new signal for the observation of $\nu_x$ neutrinos 
in water \v{C}erenkov detectors has been proposed
\cite{Langanke}.
Schematically the detection scheme works
as follows. Supernova $\nu_x$ neutrinos, with average energies
of $\approx 25$ MeV, will predominantly excite $1^-$ and $2^-$ giant
resonances in $^{16}$O via  the 
$^{16}$O($\nu_x,\nu^\prime_x$)$^{16}$O$^\star$ neutral current reaction
\cite{Kolbe2}. These resonances are above the particle thresholds and
will mainly decay by proton and neutron emission. 
Although these
decays will be dominantly to the ground states of
$^{15}$N and $^{15}$O, respectively, some of them will go to
excited states in these nuclei. If these excited states are below
the particle thresholds in 
$^{15}$N ($E^\star < 10.2$ MeV) or
$^{15}$O ($E^\star < 7.3$ MeV), 
they will decay by $\gamma$ emission.
As the first excited states in both nuclei 
($E^\star =5.27$ MeV in $^{15}$N and
$E^\star = 5.18$ MeV in $^{15}$O) 
are at energies larger than the SK detection threshold, all of the excited
states in $^{15}$N and $^{15}$O below the respective particle
thresholds will emit photons which can be observed in SK.

Based on a calculation which 
combines the
Continuum RPA with the statistical model 
\cite{Langanke}, 
Superkamiokande is expected to observe about 360
$\gamma$ events in the energy window $E=5-10$ MeV, induced by
$\nu_x$ neutrinos (with a FD distribution of $T=8$ MeV),
for a supernova going off at 10 kpc ($\approx 3 \cdot 10^4$ lightyears
or the distance to the galactic center).
This is to be compared
with a smooth background of about 270 positron events from the
 ${\bar \nu_e} + p \rightarrow n + e^+$ reaction in the same 
energy window (see Fig. 9).  
The number of
events produced by supernova $\nu_x$ neutrinos via the scheme proposed here
is larger than the total number of events 
expected from $\nu_x$-electron scattering (about 80
events \cite{Totsuka}). More importantly, the $\gamma$ signal
can be unambiguously identified from the observed spectrum in the SK detector,
in contrast to the more difficult identification from $\nu_x$-electron
scattering. 
The proposed scheme will also increase the neutral current signal for
supernova neutrinos in the heavy water detector SNO, 
but it is weaker than the one stemming
from dissociation of the deuteron.

\subsection{Observation of supernova $\gamma$ rays}

In recent years gamma-ray astronomy has grown into a 
fascinating tool to study past
nucleosynthesis \cite{Prantzos96}. The first observation of the 1.809 MeV
$\gamma$-rays stemming from the decay of $^{26}$Al \cite{Mahoney82}
has boosted the field, and in the meantime an all-sky image of current
nucleosynthesis in our galaxy has been derived from the observations
of COMPTEL \cite{Oberlack}. Noting that $\gamma$ rays penetrate the
interstellar medium without significant interaction, 
the mean life time of $^{26}$Al
($\approx 10^6$ years) is, however, too long to assign the observed 
$\gamma$ rays to a specific supernova event.
This is different for the 1.157 MeV $\gamma$ rays observed in 1994 
in the Cas A supernova remnant  \cite{Iyudin94}
stemming from the $^{44}$Ti decay chain.

Supernova models predict a small amount of $^{44}$Ti produced in
the $\alpha$-rich freeze-out in the core. With a halflife of 60 years
\cite{Goerres,Norman}, $^{44}$Ti decays to $^{44}$Sc which then decays
fast to an excited state in $^{44}$Ca whose deexcitation produces the
observed 1.157 MeV $\gamma$ rays. As the 
CAS A supernova remnant is young and close 
(the event happened between 1668-1680 at a distance of 3.4 kpc), COMPTEL
was able to measure the flux of the line rather precisely
\cite{Iyudin97} and, knowing the halflife of $^{44}$Ti,
an ejected amount 
of about $2.4 \cdot 10^{-4}$ solar masses in $^{44}$Ti
could be deduced.

Note, however, that not all of the core matter is ejected, as matter
inside the mass cut is accreted onto the compact remnant of the
supernova. The mass cut, i.e. the coordinate which separates ejected and
accreted matter, is largely unknown and is usually 
heuristically determined in model calculations. In this way, the
determination of the $^{44}$Ti yields allow to
fix the mass cut coordinate in the models, which then predict the ejected
amount of other radioactive material. For example, 
CAS A should then have ejected about 0.05 solar masses of $^{56}$Ni,
making it a rather bright and easily visible supernova. However, the
absence of historical records suggest a rather large visual extinction
at the time of the explosion, possibly caused by extra material
distributed closely to CAS A, as suggested by ROSAT studies of the X-ray
scattering halo of the supernova remnant \cite{Rosat}.

It is interesting to note that the ejected mass of $^{44}$Ti inferred
from SN1987A (a type II event) and from CAS A (a type Ib event) is
rather similar. Nevertheless a direct observation of the 1.157 MeV
$\gamma$ rays from SN1987A has not been successful yet. Improved data
are expected from the next generation of $\gamma$ ray observatories like
INTEGRAL to be launched in 2001. One of the goals will be to determine
the Galactic supernova rate.

\section{Explosive hydrogen burning}

Wallace and Woosley have pointed out that hydrogen can also be burnt
explosively  in a variety of high-temperature, high-density environments
such as type II supernovae shock waves and binary systems involving
accretion onto white dwarfs (leading to nova explosions)
or neutron stars (x-ray bursts). At temperatures higher
than those for `normal' hydrostatic CNO-cycle hydrogen burning,
extended hydrogen burning involving other cycles of nuclei such as
NeNaMg, MgAlSi or SiPS can occur at greatly accelerated rates
\cite{Champagne}.
For temperatures ranging from a few times $10^8$ K to above $10^9$ K,
and densities ranging from $10^3$ g/cm$^3$ to $10^6$ g/cm$^3$, hydrogen
burning can transmute nuclei from the CNO region to much heavier nuclei
during explosive time scales (1-100 seconds); the higher temperature and
density, the further up the chart of nuclides the rp-process (short for rapid
proton capture) can progress. The rp-process proceeds by radiative
proton captures and $\beta^+$ decays along a path close to the proton
dripline; a few $(\alpha,\gamma$) and $(\alpha$,p) reactions will occur
as well.

Under nova conditions, temperatures and densities of $4\cdot 10^8$ K and
$10^4$ g/cm$^3$ are reached. Under these conditions the (experimentally
unknown)
$^{15}$O($\alpha,\gamma$)$^{19}$Ne rate is expected to be too slow to
connect the CNO cycle with the higher hydrogen cycles; a mass flow to
heavier nuclides will only occur if neon is already available in the
accreting white dwarf (e.g. ONeMg white dwarf). 
In this case mass flow upto the sulfur region is
possible in novae \cite{vanWoermer}.
If the accreting compact object is a neutron star, 
nuclear burning is ignited at high densities ($\approx 10^6$ g/cm$^3$)
in the accreted envelope via the pp-chain, the hot CNO cycle and the
triple-$\alpha$ process. The released energy triggers a thermonuclear
runaway;
the peak temperatures 
of  $2 \cdot 10^9$ K, which can be reached before the degeneracy is
lifted, are high enough to trigger an rp-process. Depending on the time
scale available, simulations find matter flow upto $^{56}$Ni or even
further upto $^{96}$Cd \cite{Schatz97} (see Fig. 10).

\begin{figure}[htb]
\begin{center}
  \leavevmode
  \psfig{figure=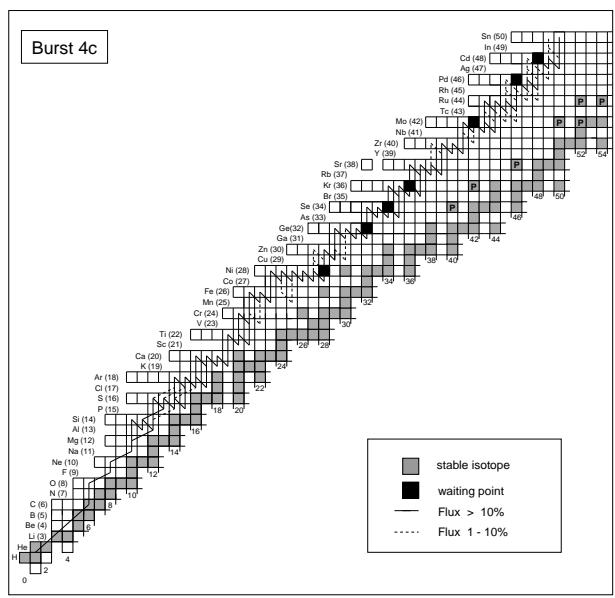,width=0.5\textwidth} 
\caption{
Reaction flow pattern in explosive hydrogen burning under x-ray burst
conditions (courtesy of H. Schatz, \protect\cite{Schatz}).
}
\label{fig:xrb}
\end{center}
\end{figure}

Similar to the r-process, studies of the rp-process require the
knowledge of masses, lifetimes and capture cross sections for unstable
(proton-rich) nuclei. Many accelerators, such as the NSCL at Michigan State
University, the TRIUMF accelerator in Vancouver, Louvain-La-Neuve in
Belgium, GSI in Darmstadt and
the RIKEN accelerator in Japan, have contributed much to the
determination of these quantities. 
It should be
noted that, unlike the r-process, the small reaction Q-values (often 
less than 2 MeV) do not permit the application of statistical model
cross sections; the relevant capture cross sections
have to be determined experimentally.
Thus, a decisive boost to simulations of
the rp-process is expected from the radioactive ion-beam facilities
currently under construction in America, Europe and Japan. 

\vspace{1.0cm}

It is a pleasure to thank R. Diehl, H.-T. Janka, G. Martinez-Pinedo
and H. Schatz to provide material used in this manuscipt. Many fruitful
discussions with F.-K. Thielemann  are gratefully
acknowledged. My own research has  benefitted from collaboration with
D.J. Dean, W.C. Haxton, E. Kolbe, S.E. Koonin, G. Martinez-Pinedo,
Y.-Z. Qian, P.B. Radha, M.R. Strayer, and P. Vogel. The research has
been partly supported by a grant of the Danish Research Council.

\end{document}